\documentclass[fleqn,usenatbib]{mnras}

\usepackage{graphicx}	
\usepackage{amsfonts,amsmath,bm,url}	
\usepackage[normalem]{ulem}  
\usepackage[utf8]{inputenc}
\usepackage[font=scriptsize,labelfont=bf]{caption}

\makeatletter
\newcommand{\setword}[2]{%
  \phantomsection
  #1\def\@currentlabel{\unexpanded{#1}}\label{#2}%
}
\makeatother

\title[Proto-neutron stars masses and radii]{What can be learned from a proto-neutron star's mass and radius?}

\author[E. Pr\'eau, A. Pascal,J. Novak and M. Oertel]{
  E. Pr\'eau,$^{1,2}$\thanks{E-mail: edwan-preau@hotmail.fr}
  A. Pascal,$^{1}$\thanks{E-mail: aurelien.pascal@obspm.fr}
  J. Novak$^{1}$\thanks{E-mail: jerome.novak@obspm.fr}
 and M. Oertel$^{1}$\thanks{E-mail: micaela.oertel@obspm.fr}\\
$^{1}$Laboratoire Univers et Th\'eories, Observatoire de Paris,
 Universit\'e PSL, CNRS, Universit\'e de Paris, 92190 Meudon, France\\
$^{2}$Astroparticule et Cosmologie, Université de Paris, CNRS, 75013
 Paris, France\\
}

\date{\today}

\pubyear{2020}

\begin{document}
\label{firstpage}
\pagerange{\pageref{firstpage}--\pageref{lastpage}}
\maketitle

\begin{abstract}
  We make extensive numerical studies of masses and radii of
  proto-neutron stars during the first second after their birth in
  core-collapse supernova events. We use a quasi-static approach for
  the computation of proto-neutron star structure, built on
  parameterized entropy and electron fraction profiles, that are then
  evolved with neutrino cooling processes. We vary the equation of
  state of nuclear matter, the proto-neutron star mass and the
  parameters of the initial profiles, to take into account our
  ignorance of the supernova progenitor properties. We show that if
  masses and radii of a proto-neutron star can be determined in the
  first second after the birth, e.g. from gravitational wave emission,
  no information could be obtained on the corresponding cold neutron
  star and therefore on the cold nuclear equation of state. Similarly,
  it seems unlikely that any property of the proto-neutron star
  equation of state (hot and not beta-equilibrated) could be
  determined either, mostly due to the lack of information on the
  entropy, or equivalently temperature, distribution in such objects.
\end{abstract}



\section{Introduction}
\label{s:intro}

Neutron stars are fascinating objects as their understanding combines
many aspects of physics under extreme conditions: relativistic
gravitational field and matter at supra-nuclear densities, to list the
most notable ones (for a more detailed description,
see e.g.~\citealt{Haensel}). They can therefore be used as astrophysical
laboratories to test our understanding of the laws of physics under
conditions which cannot be reached by Earth-based experiments. Whereas
the theory of General Relativity is tested to a high accuracy within
the neutron star context \citep{Wex2020,Kramer06}, the situation is not so
favorable when probing the properties of nuclear matter. In this
respect, one of the major objectives in the field is the determination
of the equation of state (EoS) of cold nuclear matter in
beta-equilibrium, relating e.g.~pressure $P$ and baryon number density
$n_b$. This EoS is relevant for the description of rather old (more
than a few minutes) neutron stars at the end of their chemical
evolution, i.e.~in equilibrium with respect to weak interactions and
transparent to neutrinos. Recent progress has been made with the first
detection of gravitational waves (GW) from a binary neutron star
coalescence, which has put some constraints on the neutron star tidal
deformability~\citep{LIGO18}. This first detection shows the potential
of future GW detections from binary neutron star mergers to constrain
the EoS, see e.g.~\citep{De2018, Dexheimer2018, Capano2019, Malik2019,
  Guven2020}. There exists in fact a one-to-one relation between a
given EoS and the tidal deformability as function of the star's
mass~\citep{Hinderer2010} similar to the \emph{mass-radius diagram}
(M-R diagram) of cold non-rotating neutron stars. The latter is the
subject of the standard approach for inferring constraints on the
nuclear matter EoS~\citep{Ozel2010, Steiner2010, Steiner2012,
  Ozel2015}. Therefore, many efforts are underway to measure at least
some points in this diagram, see for example recent results by
NICER~\citep{NICER19, Riley2019}.

In a prospective way, \cite{TorresForne19}, have numerically
determined empirical relations between masses ($M_{\rm PNS}$), radii
($R_{\rm PNS}$) and oscillation frequencies of proto-neutron stars
(PNS). The PNS forms during core-collapse supernova (CCSN) events,
which should be good sources of gravitational radiation, too. The
detection of GW from such modes could thus directly be related to PNS
properties. With a numerical perturbative approach in General
Relativity~\cite{TorresForne18, TorresForne19b} have identified
oscillation modes of a newly born PNS during the first second of its
life and related them to the PNS surface gravity
($M_{\rm PNS} / R_{\rm PNS}^2$) for so-called $g$-modes and shock
properties for $p$-modes. With some additional assumptions, it is then
possible to deduce $M_{\rm PNS}$ and $R_{\rm PNS}$ from observed mode
frequencies in GW. However, the structure of newly born PNSs during
the first seconds of their life is very different from that of cold
neutron stars~\citep{Prakash1996,Pons99}. In particular, properties of
nuclear matter cannot in general be cast into a simple function of one
variable as $P(n_b)$, but usually rely on a more complex 3-dimensional
relation such as $P(n_b, T, Y_e)$, with $T$, the temperature and
$Y_e=\left( n_e - n_{\bar{e}} \right) / n_b$ the electron fraction,
i.e.~the ratio between electron number density ($n_e$, minus the
positron number density $n_{\bar{e}}$) and $n_b$.

\begin{figure}[h]
\begin{center}
\includegraphics[width=8cm]{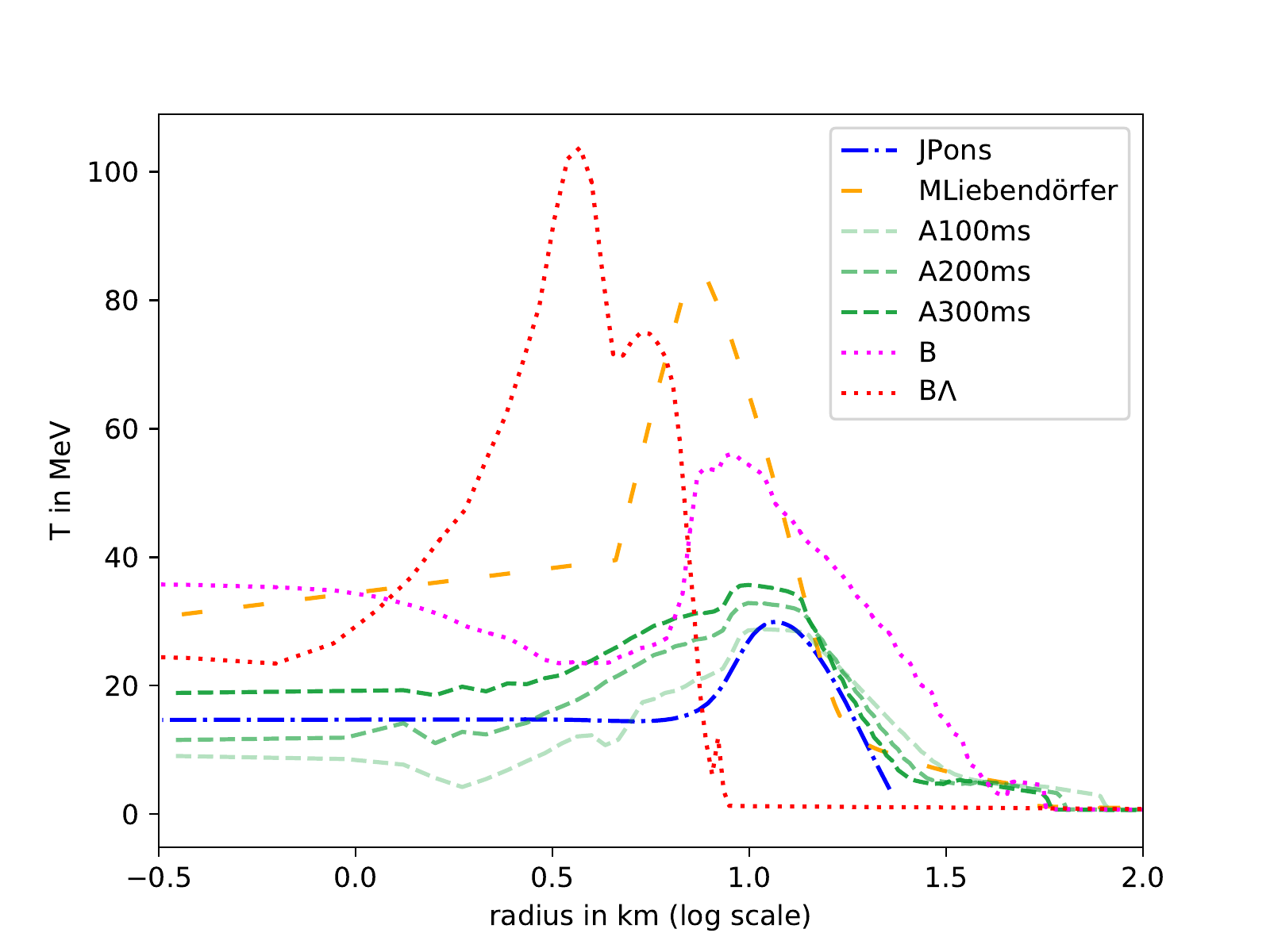}\hfill
\includegraphics[width=8cm]{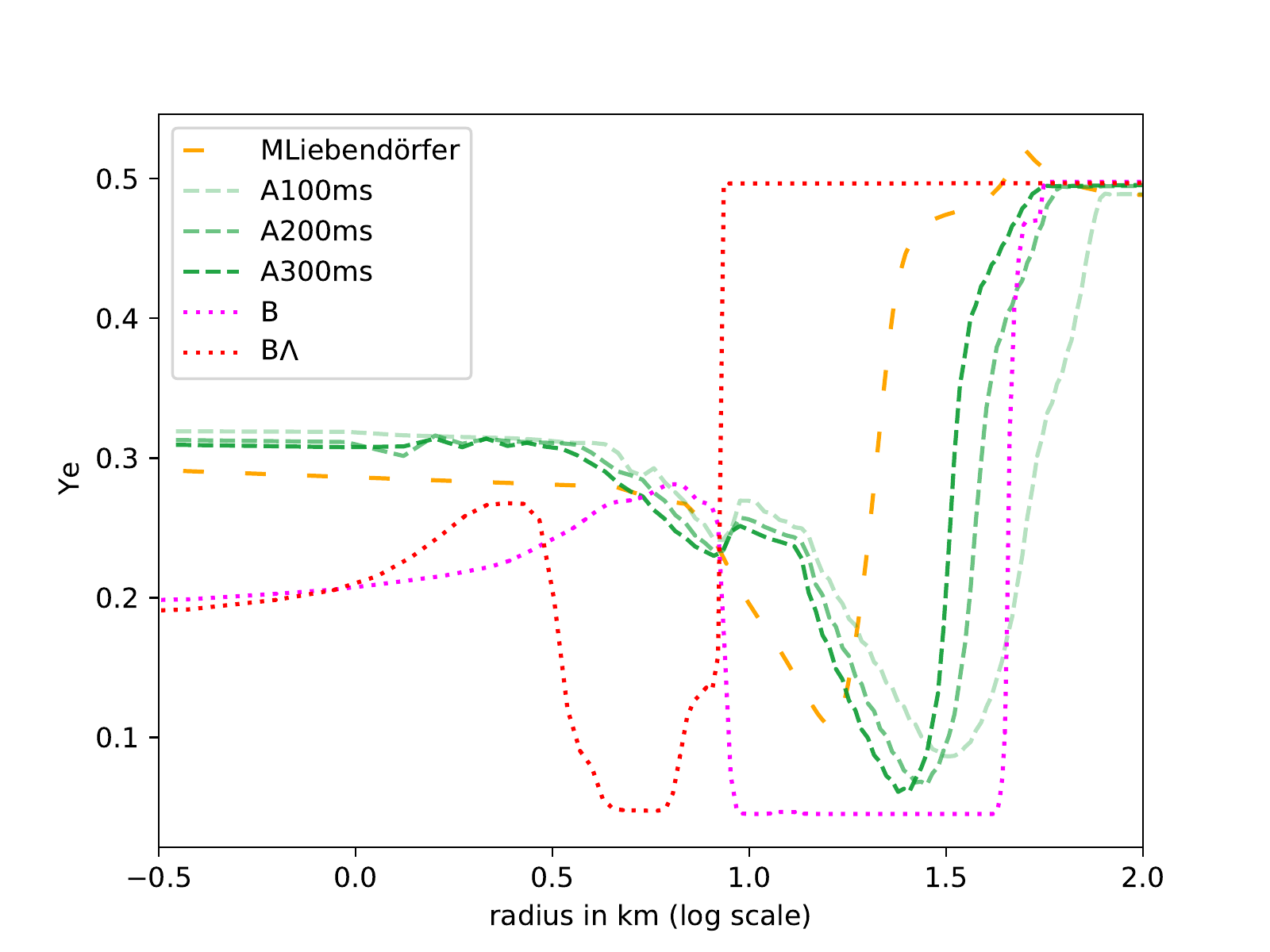}
\end{center}
\caption{Temperature $T$ (top) and electron fraction $Y_e$ (bottom)
  radial profiles obtained from various core-collapse supernova
  simulations (see text for details).}
\label{plotsT_Ye}
\end{figure}
In this context, the aim of this work is to investigate two
problems. \setword{\textbf{(P1)}}{P1} how much would the determination
of the mass and radius of a PNS born in core-collapse supernovae be
able to constrain the M-R diagram of \emph{cold} neutron stars, and
therefore also their cold equation of state $P(n_b)$ ? \\The second
one \setword{\textbf{(P2)}}{P2} is related to the first one: is it
possible, with such observations, to constrain at least the
(3-dimensional) EoS of \emph{hot} newly born PNS?

We present the model
and the methods that were used to tackle these two problems in
Sec.~\ref{s:model}. Our results are presented in Sec.~\ref{s:resus},
before some concluding remarks in Sec.~\ref{s:conc}.

\section{Quasi-static approach to proto-neutron star evolution}
\label{s:model}

In this work, we shall only consider models in spherical symmetry,
i.e. non-rotating, non-magnetized PNS, and we develop in this section
a quasi-static approach to derive their structure. Although a PNS is
an evolving star through cooling processes, the quasi-static
hypothesis is justified by the fact that the timescale necessary to
establish hydrodynamic equilibrium is much shorter than the PNS
evolution timescale, mostly determined by cooling through neutrino
emission \citep{Pons99}. Following this work, we consider models in
hydrostatic equilibrium, meaning that we follow the evolution of the
PNS driven by the neutrino cooling processes, but at each time-step,
we solve for hydrostatic equilibrium. It is in particular assumed here
that the eventual supernova explosion is successful, so that matter
outside the shock is excised from the numerical grid, once this shock
has crossed the matter of the PNS \citep[see the discussion
e.g. by][]{Roberts2016}.

The starting point is the expression for the metric of a static,
spherically symmetric spacetime, using the standard Schwarzschild-type
coordinates:
\begin{equation}
  \label{ds2} \mathrm{d}s^2 = -c^2\mathrm{d}t^2
  \mathrm{e}^{2\Phi} + \mathrm{e}^{2\lambda}\mathrm{d}r^2 +
  r^2(\mathrm{d}\theta^2 + \sin^2\theta \, \mathrm{d}\phi^2) \, ,
\end{equation}
where $\Phi(r)$ and $\lambda(r)$ are the two gravitational
potentials. It is often more convenient to introduce
a new function $m(r)$ such that:
\begin{equation}
\label{defm} \mathrm{e}^{-\lambda} = \sqrt{1 - 2Gm/(rc^2)} \, .
\end{equation}
In the Newtonian limit, $m(r)$ represents the gravitational mass
inside a sphere with radial coordinate $r$. The total gravitational
mass of the star $M_g = m(R)$ is the quantity that appears in the
Schwarzschild solution for the metric outside the star. Note that it
is this mass that could be determined through GW observation of $p$-
and $g$-modes ($M_{\rm PNS} = M_g$). Determining the structure of a
star amounts to deriving the \emph{stress-energy tensor} field
$T_{\mu\nu}$, related to the metric in Eq.~\eqref{ds2} through
Einstein Equations:
\begin{equation}
\label{EE} \mathcal{R}_{\mu\nu} - \frac{1}{2}g_{\mu\nu}\mathcal{R} =
\frac{8\pi G}{c^4} T_{\mu\nu} \, ,
\end{equation}
where $\mathcal{R}_{\mu\nu}$ is the Ricci tensor of the metric and
$\mathcal{R} = g^{\mu\nu}\mathcal{R}_{\mu\nu}$ the Ricci scalar.

\begin{figure}[h]
\begin{center}
\includegraphics[width=8cm]{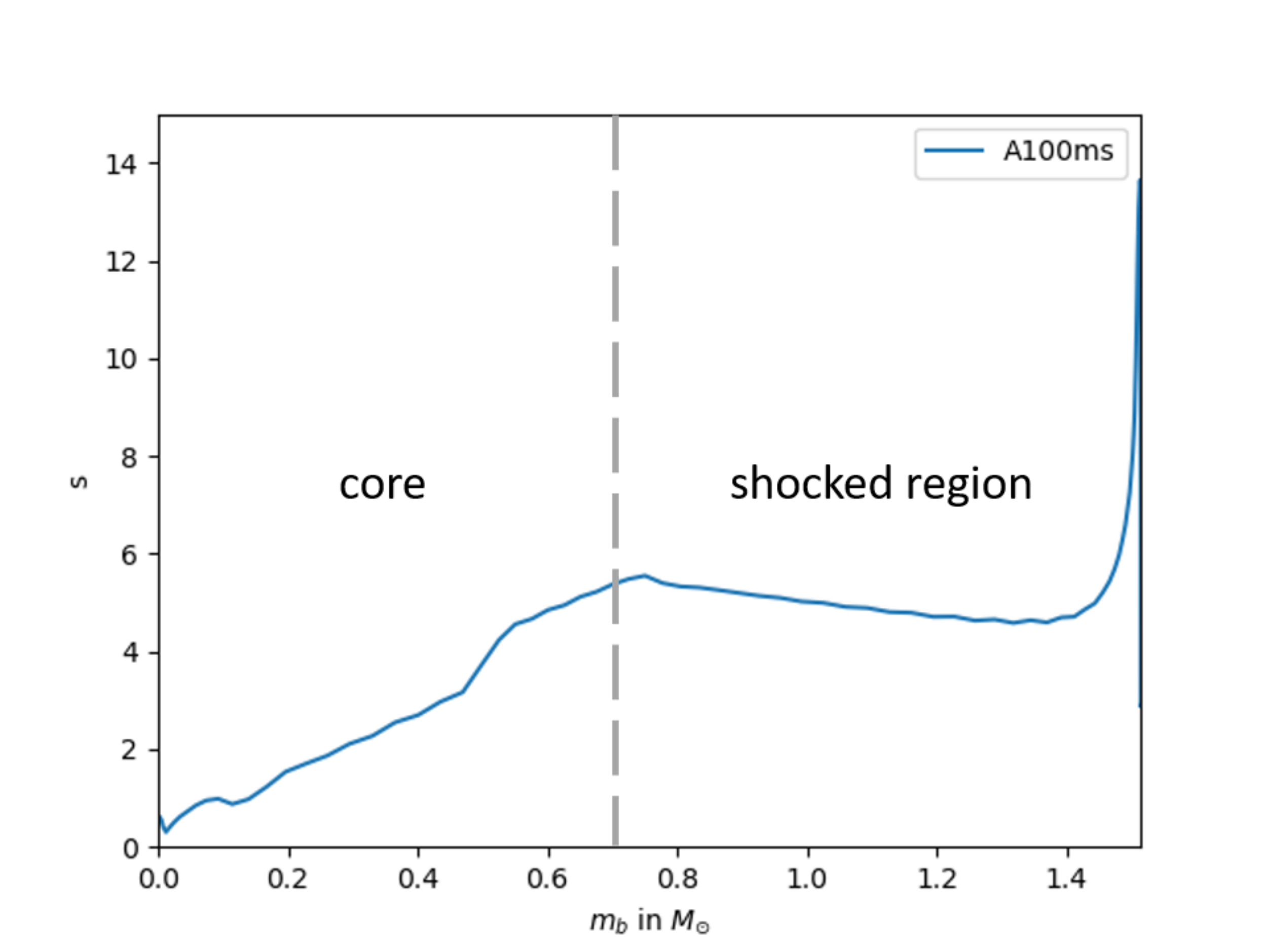}\hfill
\includegraphics[width=8cm]{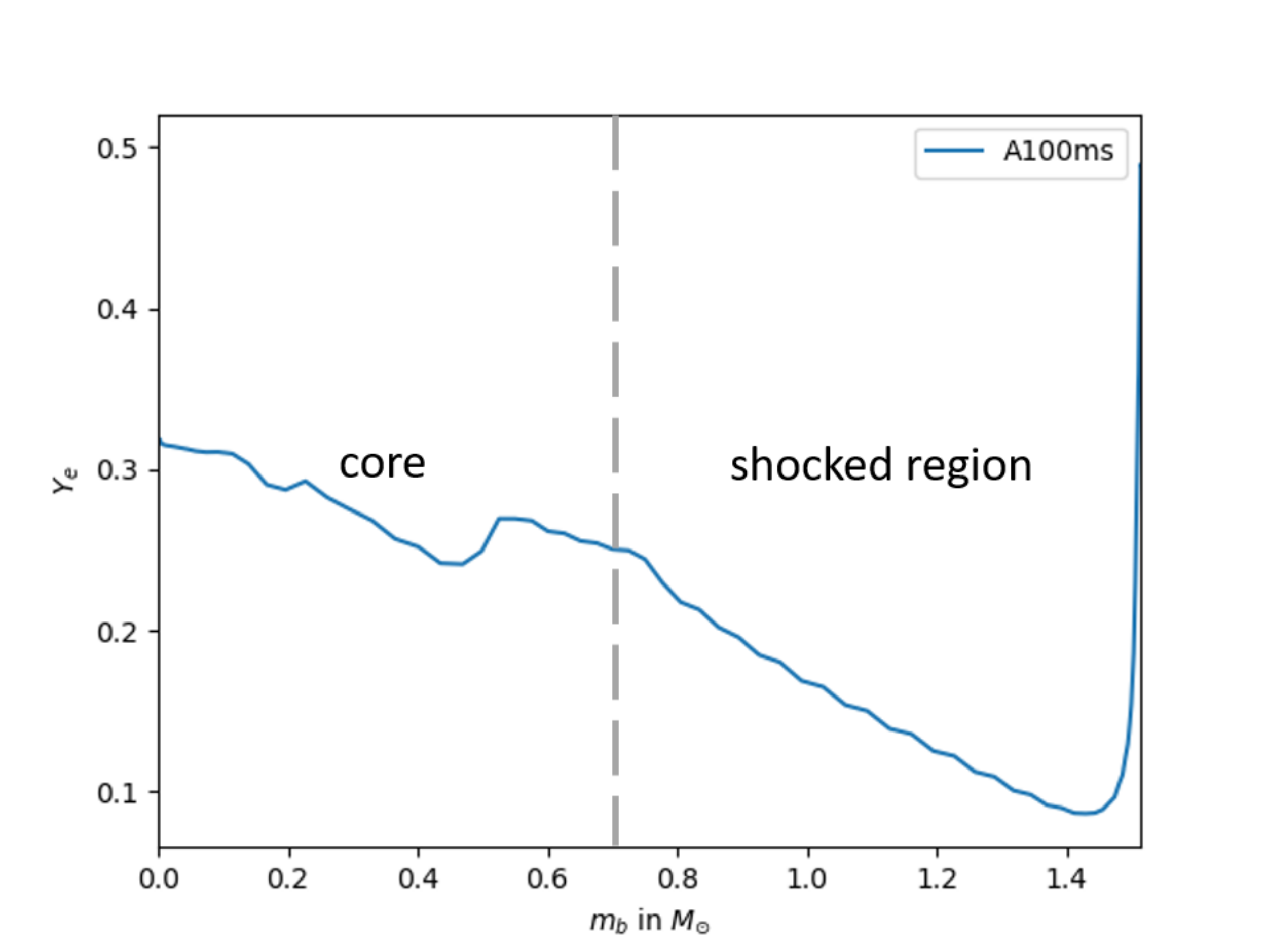}
\end{center}
\caption{Entropy per baryon $s$ (top) and electron fraction $Y_e$
  (bottom) as a function of the enclosed baryon mass $m_b$ for the
  configuration of profile A100ms from Fig.~\ref{plotsT_Ye}. The
  dashed line separates the inner core of the proto-neutron star from the
  shocked region.}
\label{sYemb}
\end{figure}
We then assume the neutron star matter to be made of a single perfect
fluid and neglect any other contribution to the stress-energy tensor:
$T_{\mu \nu} = (P + \mathcal{E})u_{\mu} u_{\nu} + P g_{\mu\nu}$ ,
where $u_{\mu}$ is the 4-velocity of matter, $\mathcal{E}$ the total
energy density and $P$ the pressure. This approximation is justified
for newly born neutron stars because shear stresses produced by strong
magnetic fields are generally negligible compared with the
pressure. On the other hand, the PNS is hot: crust is not formed yet
and superfluid effects are not present, either.

Einstein equations~\eqref{EE} in this case yield the well known
Tolman-Oppenheimer-Volkov (TOV) system for hydrostatic equilibrium for
a static spherically symmetric star:
\begin{eqnarray}
  \label{TOVm}\frac{\mathrm{d}m}{\mathrm{d}r}
  &=& \frac{4\pi}{c^2} r^2\mathcal{E}  \, , \\
  \label{TOVphi}\frac{\mathrm{d}\Phi}{\mathrm{d}r}
  &=& \frac{G m}{r^2 c^2} \left(1+\frac{4\pi
      Pr^3}{mc^2}\right)\left(1-\frac{2Gm}{c^2r}\right)^{-1}  \, , \\
    \label{TOVp}\frac{\mathrm{d}P}{\mathrm{d}r}
  &=& -\left(\mathcal{E}+P\right) \frac{\mathrm{d}\Phi}{\mathrm{d}r} \, ,
\end{eqnarray}
where Eq. \eqref{TOVp} corresponds to the projection along the
$r$-axis of the conservation of the stress-energy tensor describing
hydrostatic equilibrium. The final ingredient for the system
constituted by Eqs.~\eqref{TOVm}-\eqref{TOVp} to be closed is an EoS,
that is a relation of the form $P = P(\mathcal{E})$ (equivalent to
$P(n_b)$ introduced above). Since in a PNS thermal effects are not
negligible and weak (beta-) equilibrium is not reached, a generic EoS
for our case depends additionally, as stated in the Introduction, on
the temperature $T$ and the composition of matter summarized by the
electron fraction $Y_e$~\citep{Oertel17}. Note that in presence of
muons, the EoS would additionally depend on a muon fraction. Due to
their higher masses, they contribute, however, only marginally to the
EoS and we will neglect them here.  A generic EoS then takes the form
$P = P(\mathcal{E}, T, Y_e)$.

Consequently, in order to get an effective relation of the form $P =
P(\mathcal{E})$ and integrate the TOV system to derive the PNS
structure, it is necessary to assume some profiles $T(r)$ and $Y_e(r)$
or any equivalent ones inside the star. These profiles can be obtained
from CCSN simulations and present a high variability, depending on the
kind of progenitor that is considered for the supernova and on the
time considered after the formation of the PNS. Some examples are
plotted in Fig.~\ref{plotsT_Ye}: temperature and electron fraction
radial profiles coming from various CCSN simulations published in the
literature are shown in order to give an idea of the existing
uncertainties. To be more precise, the blue dashed-dotted curve
corresponds to the profile at $0\, \mathrm{s}$ in Fig.~12 of
\citet{Pons99}, the orange dashed curve to the model \texttt{s40} of
\citet{2004ApJS..150..263L} and \citet{2009A&A...499....1F} at $400\,
\mathrm{ms}$ after the bounce, the green dashed curves to profiles
obtained using the CoCoNuT code \citep{dimmelmeier-05,Pascal2019} for
a \texttt{s15}-type progenitor at $100\, \mathrm{ms}, 200\,
\mathrm{ms}\,\, \mathrm{and}\,\, 300\, \mathrm{ms}$ after the bounce,
the magenta dotted curves to the profile for the model \texttt{lsu40}
of \citet{peres-13} and the red dashed curve to the model
$\Lambda$-\texttt{u40} of \citet{peres-13}, both just before the
collapse of the star into a black hole.

Since such profiles can \textit{a priori} not be directly measured in
a supernova event, their variability should be taken into account when
making estimates about the structure of PNS. In other words, it is
necessary, for any given EoS, to test many realistic profiles for $T$
and $Y_e$ in order to say something about the range of possible PNS
structures that could arise for this EoS. In order to avoid performing
a large number of computationally expensive CCSN simulations, we will
use \emph{parameterized} initial profiles at the bounce to investigate
PNS structure.

These profiles are then evolved using a quasi-static evolution code (\cite{Pascal20b}): between two timesteps the electron fraction $Y_e$
and the entropy per baryon $s$ are evolved using equations
(\ref{eq:pns_evol1})-(\ref{eq:pns_evol2}) (see e.g. \cite{Pons99} and \cite{Roberts2016})
\begin{align}
    \frac{\mathrm{D}Y_e}{\mathrm{D}t} &= \frac{\mathrm{e}^{\Phi}
                                        S_n}{n_b} \label{eq:pns_evol1}
  \\
    \frac{\mathrm{D}s}{\mathrm{D}t} &=  \frac{ \mathrm{e}^{2\Phi} S_e
                                      - \mathrm{e}^{\Phi} \mu_e S_n
                                      }{n_b T} \label{eq:pns_evol2}
\end{align}
where $\mu_e$ represents the electron chemical
potential. Eq.~(\ref{eq:pns_evol1}) stems from lepton number
conservation and Eq.~(\ref{eq:pns_evol2}) from energy
conservation. For the latter, general thermodynamic relations have
been used to express energy conservation in terms of entropy per
baryon. The two source terms $S_n$ and $S_e$ are respectively the
source of electrons and the source of energy issued from
neutrino-matter interactions with a 3-flavor neutrino transport.
They are given by
\begin{align}
  S_n &= - \frac{1}{c} \left( \Gamma_{\nu_e} - \Gamma_{\bar\nu_e} \right) \\
  S_e &= - \frac{1}{c} \left( Q_{\nu_e} + Q_{\bar\nu_e} + 4Q_{\nu_x} \right)
\end{align}
with $\Gamma_{\nu_e} - \Gamma_{\bar\nu_e}$ the net electron neutrino
creation rate which, by lepton number conservation, is related to the
net change in electron number and $Q_\nu$ the energy loss rate induced
by neutrino species $\nu$. Those rates are computed using a ``Fast
Multi-group Transport" (FMT) scheme~\citep{Muller2015}. This scheme is
based on a stationary approximation of neutrino transport and uses a
two-ray approximation in areas with a high optical depth and a
two-moment closure in areas with a low optical depth. Considered reaction rates are the same as in the original paper \cite{Muller2015}.

For technical convenience, we choose here to parameterize the initial
profiles that enter the PNS evolution code as the entropy per baryon
as a function of the enclosed baryon mass $s(m_b)$ and the electron
fraction $Y_e(m_b)$. In spherical symmetry, it is equivalent to use
the radial coordinate $r$ or the enclosed baryon mass,
\begin{equation}\label{e:defmb}
m_b = 4\pi \int_0^r e^\lambda n_b m_n
r^2 \mathrm{d}r\, .
\end{equation}
The latter presents the advantage of being a Lagrangian coordinate.
Fig.~\ref{sYemb} represents typical $s(m_b)$ and $Y_e(m_b)$ profiles
obtained from CCSN simulations. Similar profiles can be found, e.g. in  Fig.~1 of \cite{2012ApJ...755..126R}.  As can be seen, these
profiles have the following characteristics: $s(m_b)$ first increases
in the inner core of the star, then slightly decreases in a region
corresponding to the shocked material and increases again very fast
towards the surface. Here, the ``inner core'' should not be confused
with the NS inner core, but be understood as the inner regions of the
progenitor star. At bounce the shock has formed at the border of the
inner core. Note that a more massive inner core (a wider region of
increasing entropy per baryon in the $s(m_b)$ profile) implies that
the kinetic energy of the shock is higher, such that it goes through
more matter before exhaustion, which will result in a larger amount of
shocked material (a wider region of decreasing entropy per baryon in
the $s(m_b)$ profile). This means that the relative mass contained in
both regions should be almost independent of the initial conditions
\citep{Janka2012a}, as in the profile of Fig.~\ref{sYemb}. Second,
$Y_e(m_b)$ is nearly constant $\sim 0.2-0.3$ in the inner core, then
decreases more sharply in the shocked region and goes up again near
the surface. We find that the following parameterizations of $s(m_b)$
and $Y_e(m_b)$ reproduce these main features.

$s(m_b)$ is modeled by a plateau followed by a Gaussian and a
power-law divergence near the surface, with five parameters in total:
\begin{eqnarray}
  \label{params} s(m_b)
  &=& s_c \times \left[ 1 +
      \exp\left(\frac{2 s_{\mathit{max}}}{s_c\sigma_s}\left(m_b -
      m_s + 2 \sigma_s/3 \right)\right) \right]^{-1}\nonumber \\
   &\hphantom{=}& +\, s_{\mathit{max}}\,\exp\left(- \frac{\left(m_b -
                           m_s\right)^2}{2\left(\sigma_s/2\right)^2}\right)\nonumber
\\
    &\hphantom{=}& +\, \frac{s_{\mathit{max}}\, \left(M_b-m_b\right)^{-0.3}}{1 +
                         \exp\left(-\frac{2}{\sigma_s}\left(m_b
                         - \left(m_s+2\sigma_s/3\right)\right)\right)} \, ,
\end{eqnarray}
where $s_c$ is the central entropy per baryon, $s_{\mathit{max}}$ the maximal
entropy par baryon between inner core and shocked material, $\sigma_s$ the
width of the Gaussian, $m_s$ the mass at which $s_{\mathit{max}}$ is
reached and $M_b$ the total baryon mass of the star. According to the
remarks above, $m_s$ and $M_b-m_s$ should be of the
same order for consistency with respect to the shock physics.
\begin{figure}[h]
\begin{center}
\includegraphics[width=8cm]{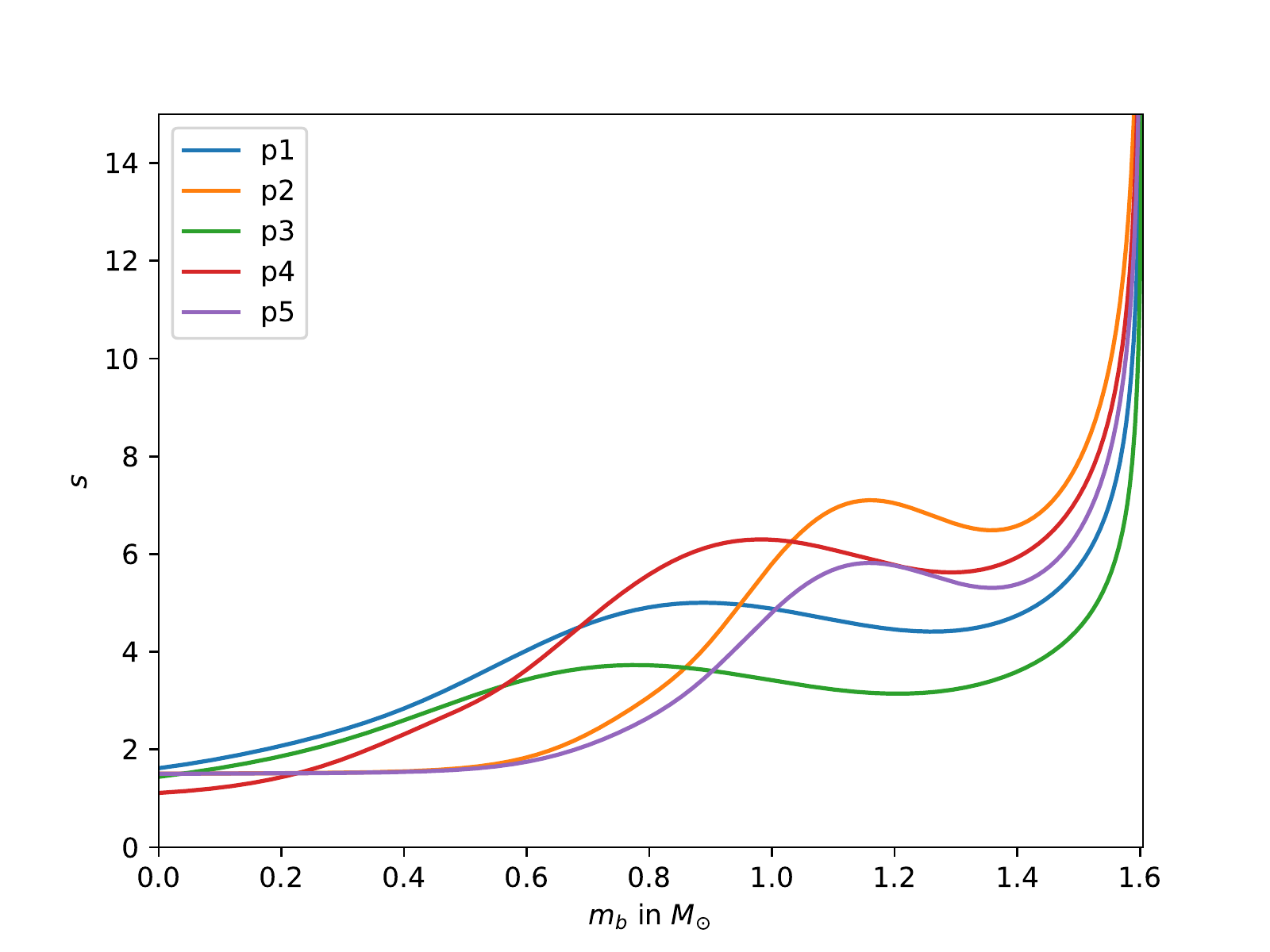}\hfill
\includegraphics[width=8cm]{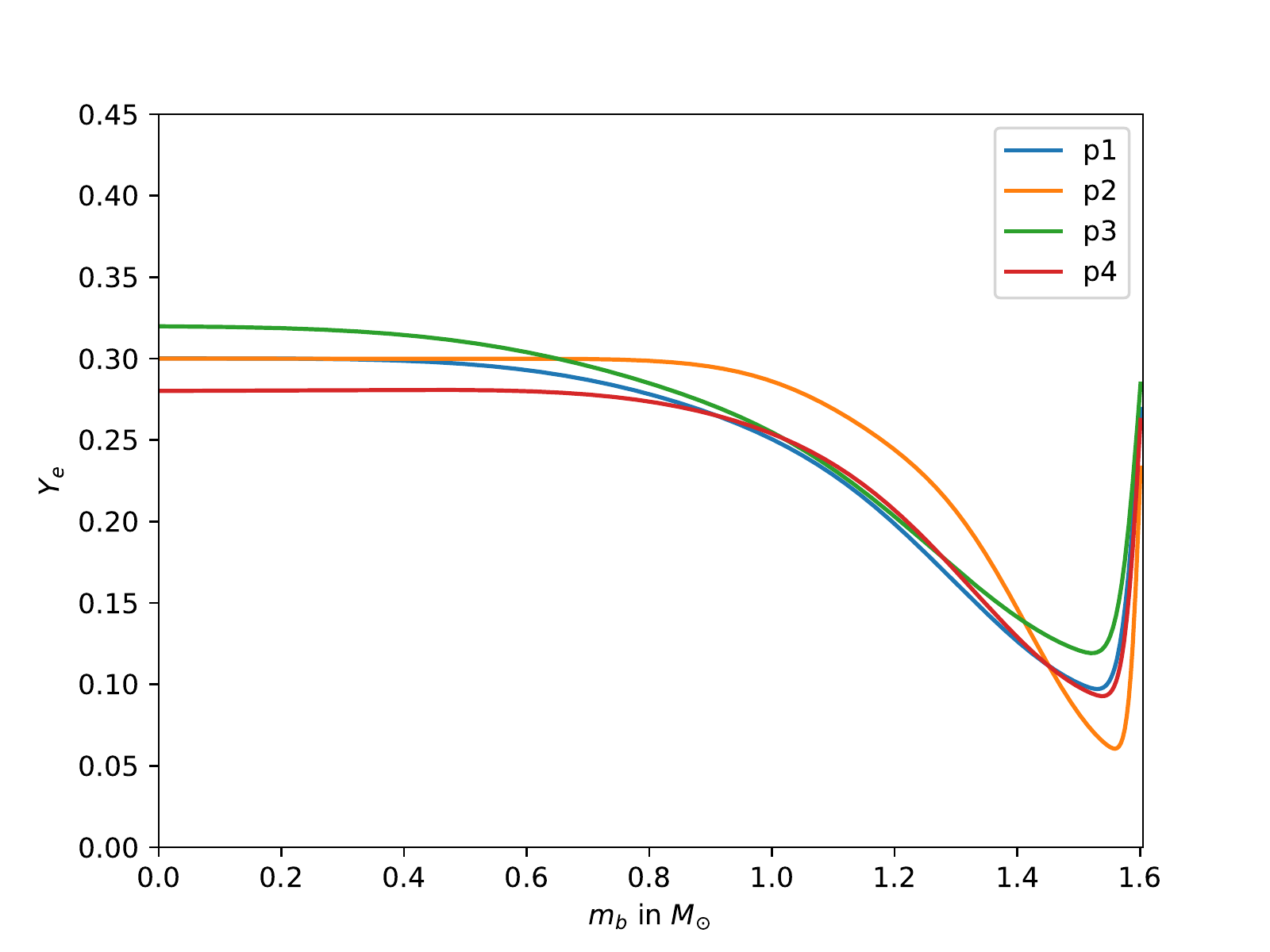}
\end{center}
\caption{Entropy per baryon $s$ (top) and electron fraction $Y_e$
  (bottom) as a function of the enclosed baryon mass $m_b$, for our
  chosen sets of parameterized
  profiles~\eqref{params}-\eqref{paramYemb} defined at bounce, for a
  total baryon mass $M_b = 1.6\, M_{\odot}$. $Y_e$ profiles for the p2
  and p5 parameterizations are identical.}
\label{sYemb_bis}
\end{figure}

$Y_e(m_b)$ is modeled by a plateau followed by an inverse
asymmetric Gaussian, with four parameters in total:
\begin{eqnarray}
  \label{paramYemb} \nonumber
  Y_e(m_b) &=& Y_0 \nonumber \\ & & + \frac{Y_c - Y_0}{1 +
               \exp\left(\frac{5}{2 \sigma_Y} \frac{Y_{\mathit{min}} - Y_0}{Y_c -
               Y_0}\left(m_b - M_b +
               \frac{2\sigma_Y}{3}\right)\right)} \nonumber \\
           &\hphantom{=}& +\, (Y_{\mathit{min}} - Y_0)\, \exp\left(
                          -\frac{\left(m_b -
                          M_b\right)^2}{2\left(\sigma_Y/2\right)^2}\right)
  \nonumber\\ &\hphantom{=}& \quad \times \frac{1}{2}\left(1 +
                           \mathrm{erf}\left(-15\,\frac{m_b -
                           M_b}{\sigma_Y/\sqrt{2}}\right)\right)\,  ,
\end{eqnarray}
where ``erf'' refers to the error function and $Y_c$ is the central
electron fraction, $Y_0 = 0.5$, $Y_{\mathit{min}}$ the minimal $Y_e$
and $\sigma_Y$ the width of the Gaussian. Since the region where a
depletion in $Y_e$ is observed, here described by the Gaussian,
corresponds to the shocked region (see Fig.~\ref{sYemb}), $\sigma_Y$
should be of the same order as $M_b - m_s$.
\begin{figure}[h]
\begin{center}
\includegraphics[width=8cm]{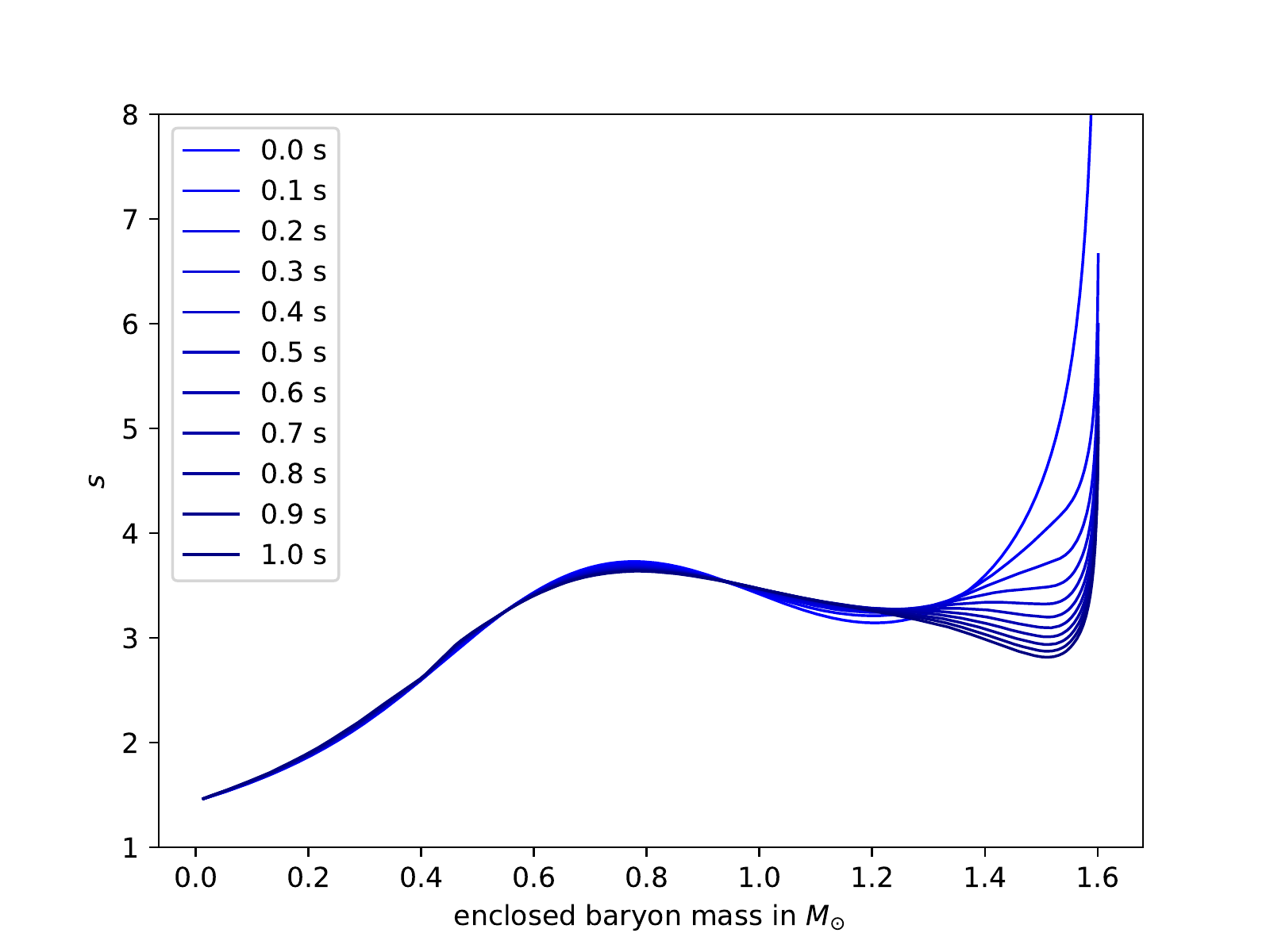}\hfill
\includegraphics[width=8cm]{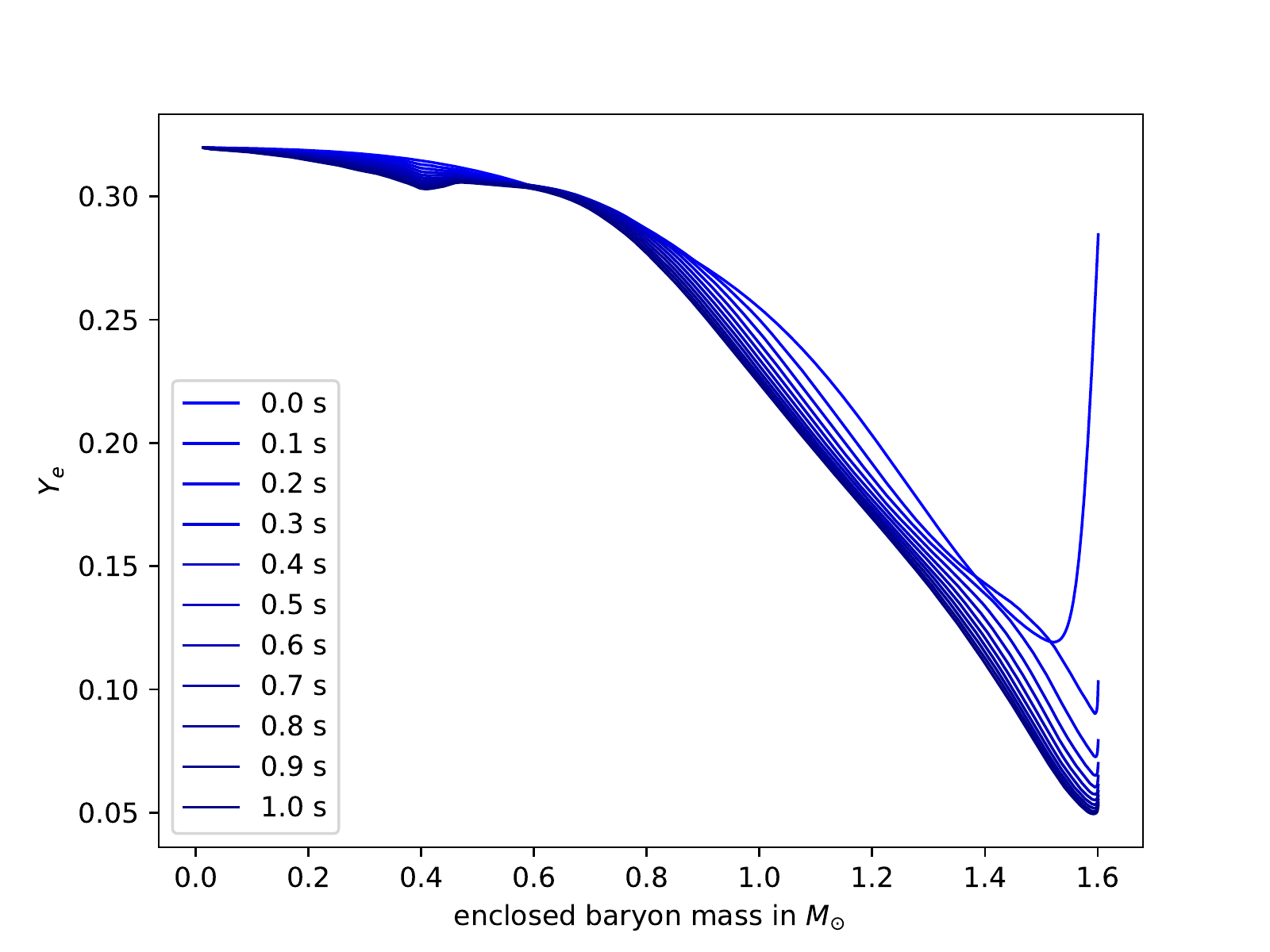}\hfill
\includegraphics[width=8cm]{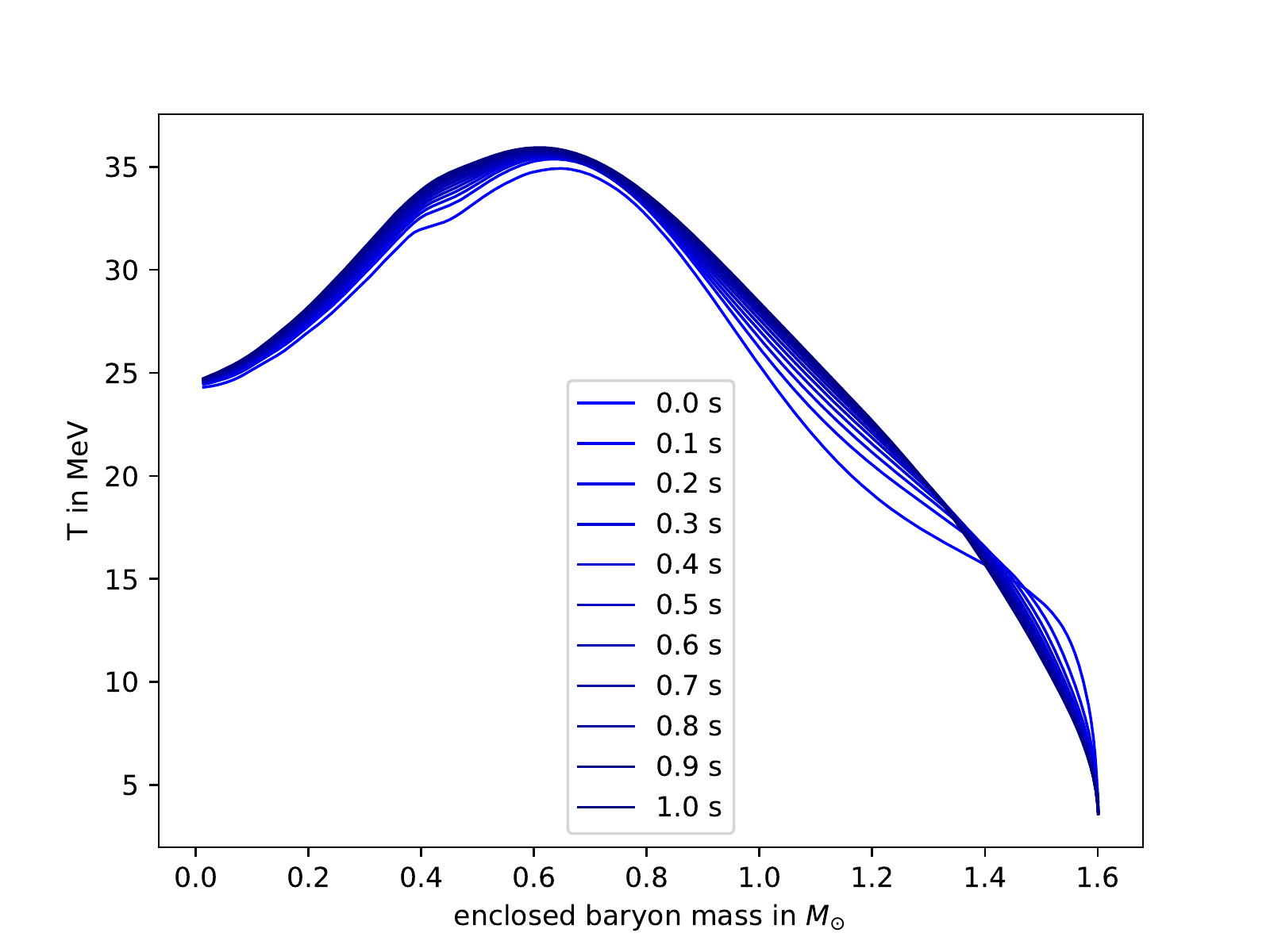}
\end{center}
\caption{Entropy per baryon $s(m_b)$ (top), electron fraction
  $Y_e(m_b)$ (middle) and temperature $T(m_b)$ (bottom) profiles as a
  function of enclosed baryon mass $m_b$, for several post-bounce
  times (darker blue corresponds to later times) employing the initial
  profile p3 of Tab.~\ref{tab:param}, total baryon mass $M_b = 1.6\,
  M_{\odot}$ and the HS(DD2) EoS~\citep{Hempel_NPA_2010}. The time
  evolution is computed within a quasi-static approach, see
  Sec.~\ref{s:model}.}
\label{sYeTmb}
\end{figure}

\section{Numerical model parameters and results}
\label{s:resus}

We present in this section the results of a comparison between several
simulations of PNS structure during the first second of their
post-bounce evolution varying the initial entropy and electron
fraction profiles as well as the employed EoS model. Let us start with
detailing the procedure applied for that comparison.

\begin{table}
\begin{tabular}{|l|c|c|c|c|c|c|p{1.5cm}|}
  \hline
Label & $s_c$  & $s_{\mathit{max}}$ & $\sigma_s$ & $m_s/M_b$ & $Y_c$ & $Y_{\mathit{min}}$  \\
  \hline
&&&&&&\\
p1 & 1.3  & 4 & 0.7 & 0.5 & 0.26 & 0.1  \\
&&&&&&\\
  \hline
  &&&&&&\\
p2 & 1.5 & 5.5 & 0.4 & 0.6875 & 0.3 & 0.06  \\
&&&&&&\\
  \hline
&&&&&&\\
p3 & 1.1 & 3 & 0.7 & 0.4375 & 0.32 & 0.12 \\
&&&&&&\\
\hline
  &&&&&&\\
p4 & 1 & 5 & 0.6 & 0.5625 & 0.28 & 0.1  \\
&&&&&&\\
  \hline
  &&&&&&\\
p5 & 1.5 & 4.5 & 0.4 & 0.6875 & 0.3 & 0.06   \\
&&&&&&\\
\hline
\end{tabular}
\caption{Parameters defining for a given $M_b$ the five parameterizations labeled p1 to p5 of the initial $s(m_b)$ and $Y_e(m_b)$ profiles.}
  \label{tab:param}
\end{table}
\begin{figure}[h]
\begin{center}
\includegraphics[width=8cm]{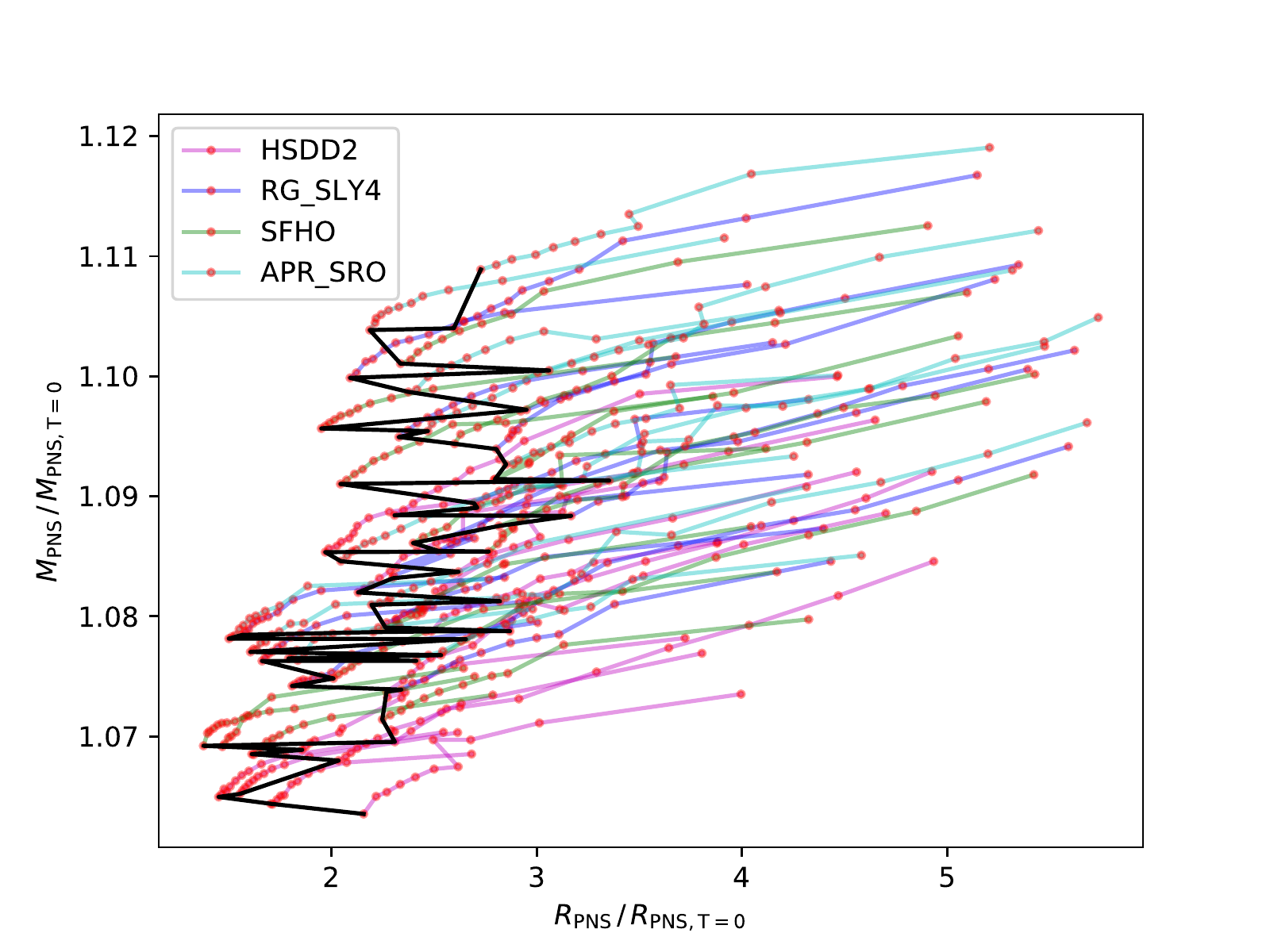}\hfill
\includegraphics[width=8cm]{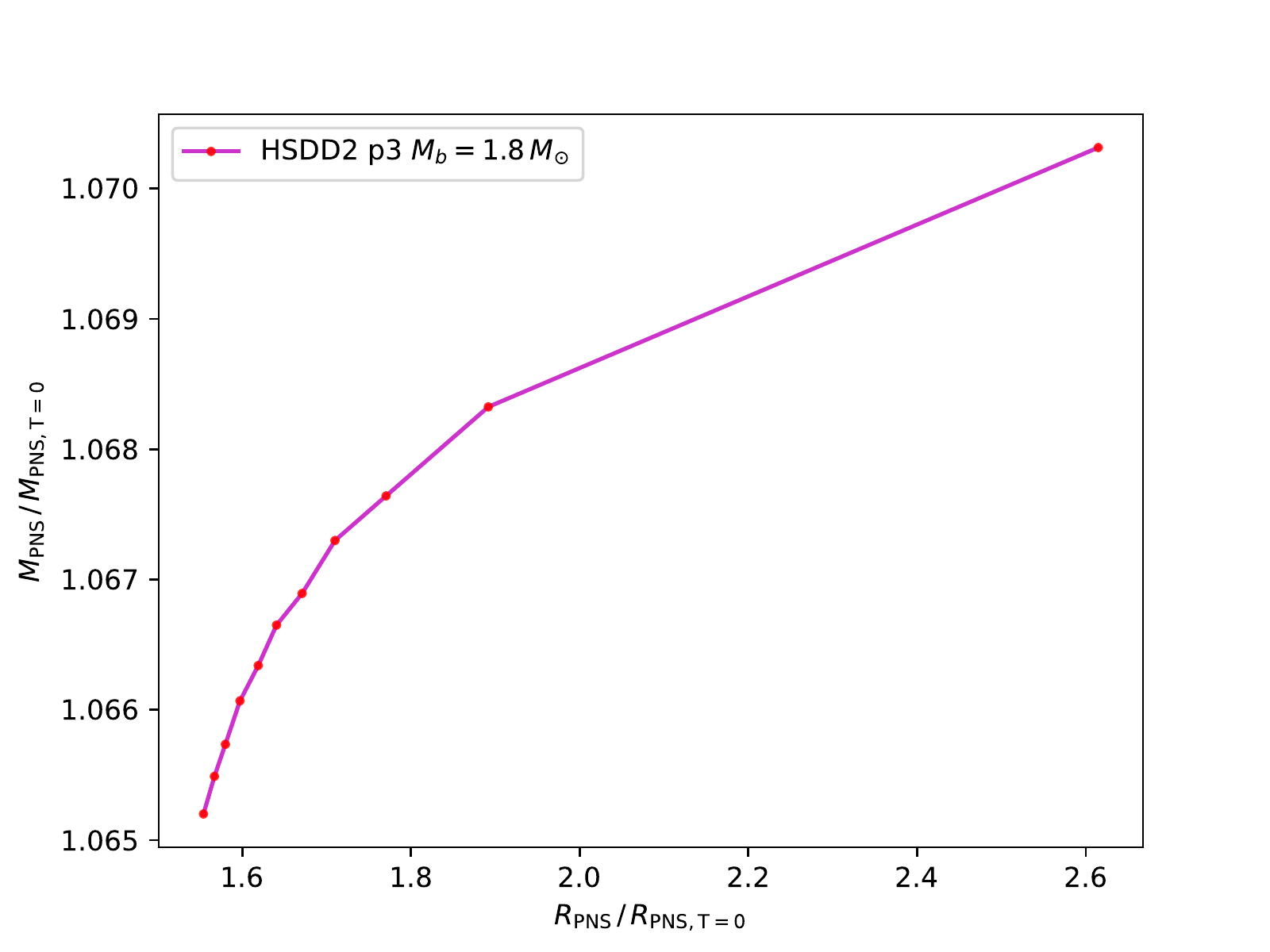}
\end{center}
\caption{Top : PNS evolutions in the gravitational
  mass~($M_{\rm PNS}$)-radius~($R_{\rm PNS}$) plane, normalized to the
  zero temperature values (subscript $T=0$), for the parameter sets p1
  - p5 of Table~\ref{tab:param}. Different colors correspond to
  different EoS models: HS(DD2)~\citep[magenta]{Hempel_NPA_2010},
  RG(SLy4)~\citep[blue]{Raduta_2019}, SFHo~\citep[green]{Steiner2013}
  and SRO(APR)~\citep[cyan]{Schneider19}. Time runs from 0~s to 1~s
  post-bounce from right to left. The red dots indicate the points
  computed every 0.1~s and the black line the points for the PNS 1~s
  after the bounce. Bottom : same curve in the particular case of the
  HS(DD2) EoS, parameterization p3 of Tab.~\ref{tab:param} and a total
  baryon mass $M_b = 1.8\, M_{\odot}$.}
\label{fig:P1}
\end{figure}

The initial profiles for $s(m_b)$ and $Y_e(m_b)$ at bounce are built
upon the parameterizations discussed in the previous section, see
Eqs.~\eqref{params}-\eqref{paramYemb}. The parameters are chosen such
that the resulting profiles remain in reasonable agreement with
results from CCSN. In particular, in order to respect the core-shock
picture discussed above and illustrated in Fig.~\ref{sYemb}, $m_s$ and
$\sigma_Y$ should be of the same order as $M_b-m_s$. For simplicity,
we choose $\sigma_Y = M_b - m_s$.  Note that $M_b$ is a parameter of
both profiles, $s(m_b)$ and $Y_e(m_b)$, so that different baryon
masses correspond to different profiles. We say that two profiles obey
the same parameterization if they have the same parameters, except for
$M_b$ which can be chosen freely. $m_s$ is then given by the ratio
$m_s/M_b \approx 1/2$. The values of the remaining free parameters for
the five parameterizations considered here are detailed in
Tab.~\ref{tab:param}. For each parameterization, three different total
baryon masses $M_b$ will be considered: $1.6\, M_{\odot}$, $1.8\,
M_{\odot}$ and $2.0\, M_{\odot}$. For illustration, the parameterized
profiles are plotted in Fig.~\ref{sYemb_bis} for a total baryon mass
$M_b = 1.6\, M_{\odot}$, the upper panel showing $s(m_b)$ and the lower
one $Y_e(m_b)$. Profile p1 is thereby chosen to resemble the profiles
of Fig.~\ref{sYemb} and p2 those of Fig.~1 in
\citet{2012ApJ...755..126R}.

\begin{figure}[h]
\begin{center}
\includegraphics[scale=0.5]{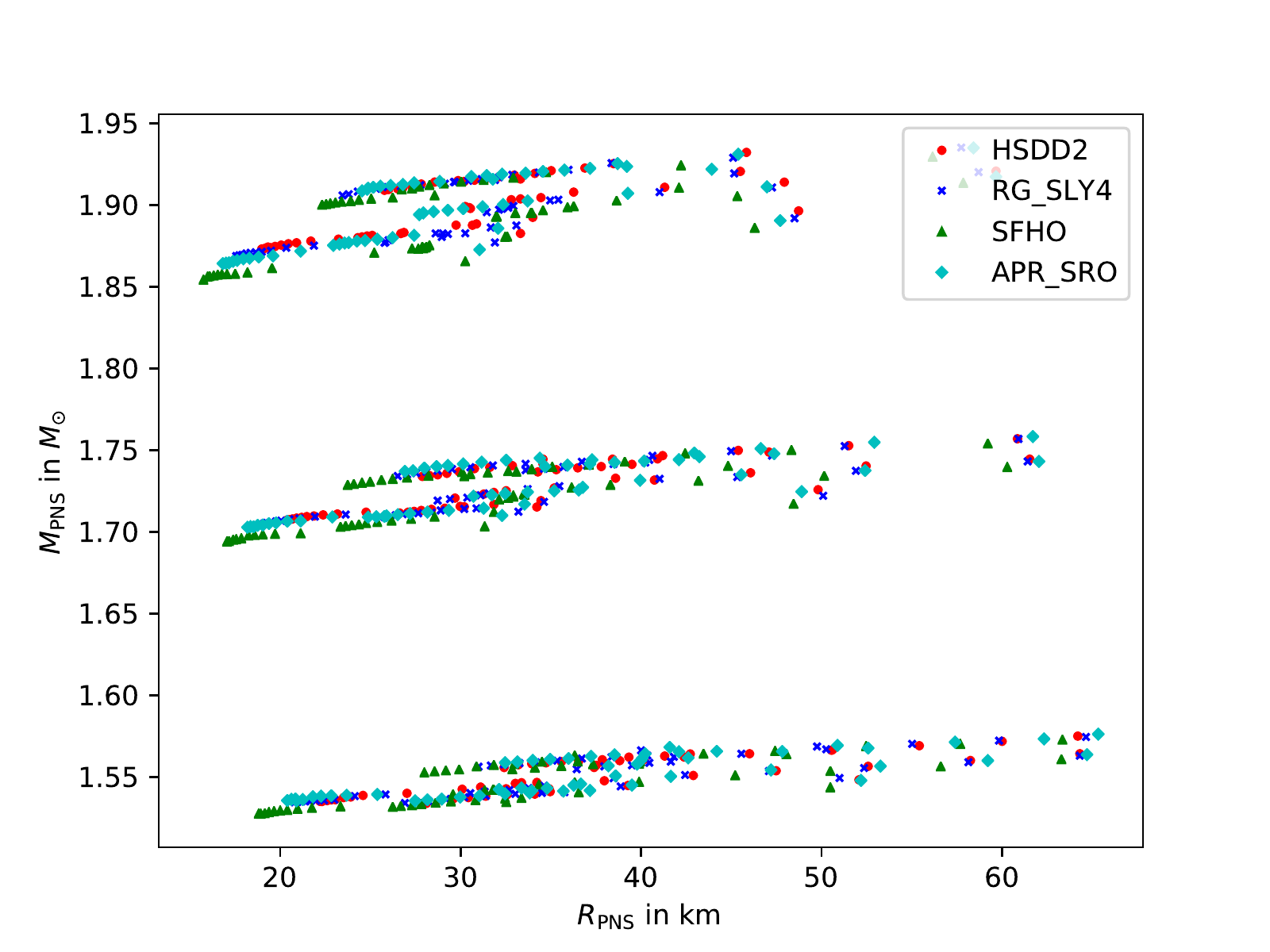}
\end{center}
\caption{Points in the $(R_{\rm PNS},M_{\rm PNS})$-plane obtained from
  the evolution of the five initial parameterizations of
  Tab.~(\ref{tab:param}), for three different total baryon masses,
  $M_b = 1.6\, M_{\odot}$ (bottom cloud), $M_b=1.8\, M_{\odot}$
  (middle cloud) and $M_b=2\, M_{\odot}$ (top cloud). 4 different EoSs
  are used: HS(DD2)~\citep[red circles]{Hempel_NPA_2010},
  RG(SLy4)~\citep[blue crosses]{Raduta_2019}, SFHo~\citep[green
    triangles]{Steiner2013} and SRO(APR)~\citep[cyan
    diamonds]{Schneider19}. }
\label{clouds}
\end{figure}
Then, with five parameterizations and three baryon masses, we end up
with a total of 15 different initial profiles. These profiles are
evolved during the first second post-bounce for a given EoS using the
quasi-static PNS evolution code~\citep{Pascal20b}, see
Section~\ref{s:model}. Fig.~\ref{sYeTmb} shows as an example the $s(m_b)$,
$Y_e(m_b)$ and $T(m_b)$ profiles in the PNS every 0.1~s post-bounce,
for the parameterization p3 of Tab.~\ref{tab:param}, a total baryon
mass $M_b = 1.6\, M_{\odot}$ and the HS(DD2)
EoS~\citep{Hempel_NPA_2010}.

Finally, with evolved profiles taken every $0.1 \mathrm{s}$
post-bounce, we end up with 150 PNS profiles for each EoS. These
profiles are used to solve the TOV system~\eqref{TOVm}-\eqref{TOVp}
and to derive the PNS structure at fixed baryon mass at each time. The
surface of the PNS is thereby defined as in \cite{TorresForne19} at a
fixed baryon density of $n_b = 5.970\times 10^{-5}\, \mathrm{fm}^{-3}$. We obtain in
the end one point in the $(R_{\rm PNS}, M_{\rm PNS})$ plane for each
of the 150 PNS profiles.

Based on the results of the procedure described above, we first
address the problem \ref{P1}, concerning the possibility, from the
observation of mass and radius of a hot PNS in the first second
post-bounce, to constrain the corresponding quantities of a
\emph{cold} neutron star. To take into account the uncertainty on the
entropy and electron fraction in the PNS, the PNS structure
is computed for the five parameterizations of initial profiles given in
Tab.~\ref{tab:param}. In Fig.~\ref{fig:P1} are displayed the time
evolution of 60 different PNS: our 15 different initial profiles with
four different choices of EoS models. These are
HS(DD2)~\citep[magenta]{Hempel_NPA_2010},
RG(SLy4)~\citep[blue]{Raduta_2019},
HS(SFHo)~\citep[green]{Steiner2013} and
SRO(APR)~\citep[cyan]{Schneider19}. The evolution is described in terms of the
radius $R_{\rm PNS}$ and the mass $M_{\rm PNS}$, each one
divided by the corresponding value for the cold configuration (label
$T=0$), corresponding to a fully-evolved (cold, beta-equilibrated) NS.
\begin{figure}[h]
\begin{center}
\includegraphics[scale=0.5]{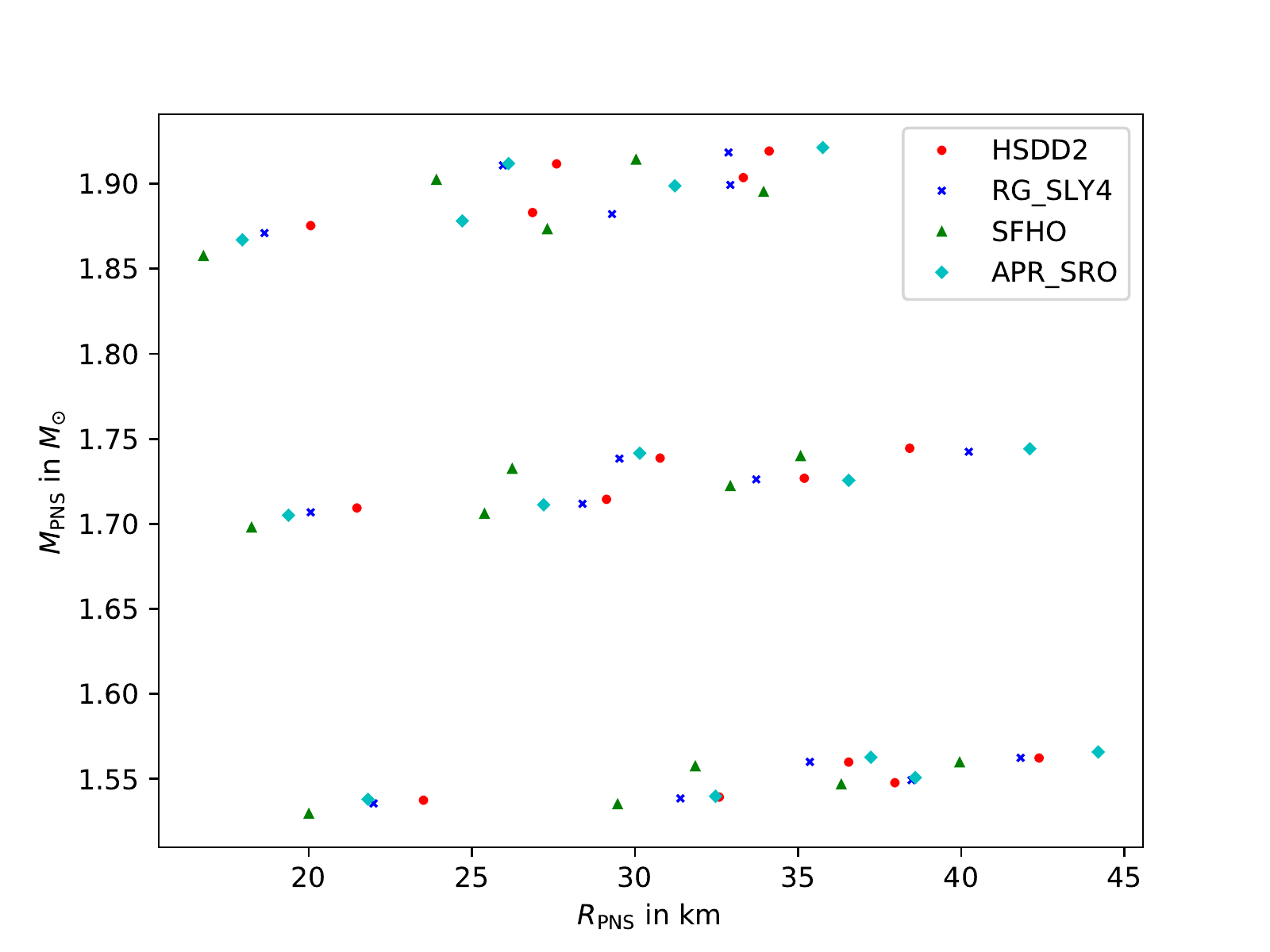}
\includegraphics[scale=0.5]{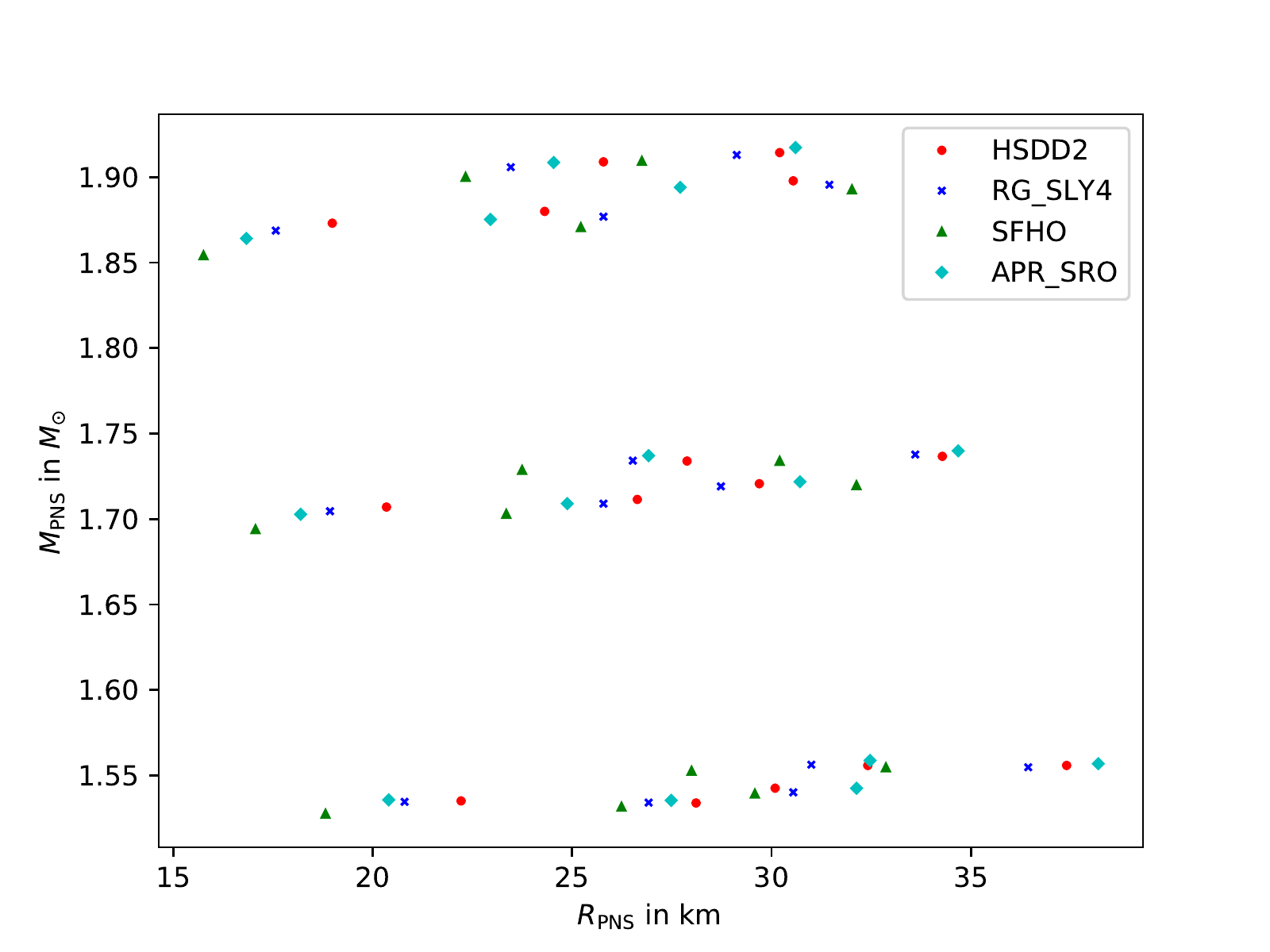}
\end{center}
\caption{Same as Fig.~\ref{clouds}, but only snapshots at
  $t=0.5$~s (top) and $t=1$~s (bottom) are taken from the time evolution.}
\label{f:time_clouds}
\end{figure}

It is apparent that for all parameterizations and EoSs, the PNS
structure is still quite far from the cold configuration after 1~s of
evolution. Quantitatively, we find that, after 1~s, the mass is larger
by about $6-10$~\% and the radius by about $40-200$~\% compared with
the cold configuration. Note that, as far as the practical observation
is concerned, because the mass is almost constant\footnote{Except in
  the first tenth of a second, when the mass should increase if
  accretion is properly taken into account} during the first second of
the PNS' life, its value is very little influenced by the time over
which the observation is integrated, so that a good signal to noise
ratio (SNR) is reachable in principle. On the contrary, integrating
over a few tenths of a second would induce a large error on the radius
measurement. All in all, our results imply that whereas measuring the
mass of a PNS during the first second post-bounce would give an upper
bound quite close to the actual value for the cold star, the obtained
value of the radius is much less reliable. Indeed during the first few
seconds of evolution of the PNS the shocked mantle cools down and
shrinks very fast~\cite{Prakash1996,Pons99}, such that the radius is
evolving on a short timescale. As a consequence, the determination of
$M_{\rm PNS}/R_{\rm PNS}^2$ proposed by \cite{TorresForne19} is
\emph{not expected to give a reliable estimate of the same quantity in
  the cold configuration}.

We now turn to the problem~\ref{P2}, regarding the possibility, from
the knowledge of $R_{\rm PNS}$ and $M_{\rm PNS}$, to constrain its
EoS, which is fully dependent on three parameters. Although the answer
would be yes if we were able to know, for each measured point at a
given post bounce time, both $s$ and $Y_e$ profiles inside the PNS,
the fact that they are not accessible implies that the actual answer
depends on the influence of our limited knowledge of these profiles on
$M_{\rm PNS}$ and $R_{\rm PNS}$. If the dispersion in
$(R_{\rm PNS},M_{\rm PNS})$ due to the uncertainty in the profiles is
smaller than the variation due to the difference induced by employing
different EoS, then this measurement can in principle (provided a
sufficient precision) constrain the EoS of the PNS. Fig.~\ref{clouds}
shows the points obtained from the PNS evolution procedure for the
four different EoS models considered before. In this figure, the
condition for the $(R_{\rm PNS}, M_{\rm PNS})$ measurement being able
to discriminate between different EoS is that the clouds of points
(corresponding to the various parameterizations for the initial $s$ and
$Y_e$ profiles) corresponding to the different EoS models do not
intersect. Actually, employing only four different EoS models, a
positive answer to problem~\ref{P2} would actually be difficult to
justify in this way. However, the clouds are largely mixed already for
a small number of different EoS. We conclude that, measuring
$R_{\rm PNS}$ and $M_{\rm PNS}$ alone can hardly constrain the EoS of
the PNS. It is \textit{a priori} possible to add some timing
information to this picture, with the measure of the mass and radius
at some given post-bounce times. To illustrate this purpose,
Figs.~\ref{f:time_clouds} show snapshots of Fig.~\ref{clouds} at
$0.5$~s and $1$~s after bounce. Here again, it is clear that the
spreading of points (mainly along the $R_{\rm PNS}$ direction) due to
the use of different entropy or electron fraction profiles is much
larger than that coming from the various EoS models. This means in
particular that, even if we can measure mass and radius of the PNS
with additional timing information, little could be deduced about the
underlying EoS.

In this context, it it is interesting to ask which of the uncertainty
on entropy or on electron fraction has the largest influence on the
large spreading in the $(M_{\rm PNS}, R_{\rm PNS})$-plane responsible
for the negative answer to \ref{P1} and \ref{P2}, or whether they both
contribute equally. To answer that question, we have introduced four
additional parameterizations based on p1: two with a modified entropy
keeping the electron fraction profile the same, and two with modified
electron fraction and the same entropy profile. For the former, we
vary $s_c$ and $s_{\mathit{max}}$ by $\pm_{20}^{50}~\% $ and
$\pm 50~\% $, respectively, compared with the p1 entropy
profile. These variations remain still in reasonable agreement with
results from simulations. For the latter, we use two rather extreme
profiles, one with a $Y_c$ and a $Y_{\mathit{min}}$ increased by
roughly 15~\% and 50~\% respectively and another one reducing $Y_c$
and $Y_{\mathit{min}}$ by 30~\% and 70~\%, respectively. Repeating the
same exercise as before, i.e. following the evolution of the PNS over
1~s with these additional initial profiles and the same four different
EoS models and analysing the obtained distribution in $M_{\rm PNS}$
and $R_{\rm PNS}$, the conclusion is rather obvious. Varying the
$Y_e$-profile, even in an extreme way, we observe only a limited
dispersion of points. As far as $M_{\rm PNS}$ is concerned, it is less
than a few percent and for $R_{\rm PNS}$, of the order of 10~\%. On
the contrary, the different entropy profiles induce a large
uncertainty, in particular for the radii which spread between roughly
15 and 60 km. Taking snapshots at given post-bounce times does not
alter the conclusion that mainly the uncertainty in the entropy
profile of the PNS prevents us from obtaining information about the
EoS and cold neutron star properties from the observation of the PNS'
mass and radius. This finding can be understood from the fact that
small variations in entropy per baryon induce large variations in
pressure and energy density of matter in particular at lower
densities, i.e. in the outer regions of the star, and that radii are
very sensitive to these variations. That means, if we want to improve
the situation and be able to extract information from the measurement
of PNS mass and radii, we should better constrain the entropy profiles
in the PNS.

It should be mentioned here, too, that the PNS is very likely unstable
to convection, flattening the entropy per baryon profile in the
shocked region \citep[see e.g. the discussion in][]{Roberts2016}. This
flattening is not reproduced by our profiles and the simulated PNS
evolution does not treat convection. Including the latter would
certainly change our results quantitatively, but the main result of
our study remains true: without considerably reducing the uncertainty
in entropy profiles, no useful information on EoS and cold NS mass and
radius can be obtained from the corresponding quantities of a PNS.

\section{Summary and discussion}
\label{s:conc}

As it seems to be possible to infer from future GW observations the
mass and the radius of a newly born PNS, we have investigated here
what information these data could bring about the nuclear EoS for
NSs (one-parameter EoS) or for PNSs (3-parameter EoS). Thermal and
composition effects cannot be neglected within a PNS. In addition,
temperature and electron fraction in a PNS evolve with time and its
mass and radius accordingly.  We have thus considered neutrino cooling
of the PNS using a quasi-static approach~\citep{Pascal20b} and
determined entropy and electron fraction profiles in the PNS at
several instants during the first second of post-bounce
evolution. Based on these profiles, we have built hydrostatic PNS
models at each step giving mass and radius. From CCSN simulations it
appears that large uncertainties exist as far as entropy and electron
fraction profiles at early post-bounce are concerned. These
uncertainties result, among others, from different progenitor
structure, uncertainties on weak reactions during infall, the employed
EoS and convection in the newborn PNS. In order to take these
uncertainties into account, we have set up analytic parameterized
profiles for both quantities ($s$ ans $Y_e$), that can mimic most of the
simulated profiles from the literature, enabling us to generate a
large set of physically relevant evolutionary models of PNSs, in the
first second after the bounce.

The study of this set of models showed that, after 1~s of evolution,
the value of the radius is still far from the corresponding value of a
cold NS. This indicates that the knowledge of PNS properties soon
after the bounce cannot give much information on the properties of the
resulting cold NS (question~\ref{P1}). Additionally, a large dispersion
of evolutionary curves of PNS in the $M_{\rm PNS}, R_{\rm PNS}$ plane
and, above all, a mixing of points coming from various EoSs is
observed. Conversely, this means that, even with some timing
information (as illustrated by Figs.~\ref{clouds} and
\ref{f:time_clouds}), it is difficult to disentangle evolutionary
tracks of models with different EoSs. This also leads to a negative
answer to the question~\ref{P2}, about the possibility of inferring
the EoS of the PNS from its masses and radii.

Of course, there are several limitations to this study. First, the
quasi-static approach may not be the most accurate, but we believe
that, as discussed in previous works~\cite{Pons99, Villain04}, such an
approximation is under control and is not the main source of
uncertainties in our results. Another point which merits discussion is
the choice of parameterized profiles. The choices done here, with
formulas in Eqs.~\eqref{params}-\eqref{paramYemb}, may be regarded as
too broad, giving too much freedom in the dispersion of points in
Figs.~\ref{fig:P1} and \ref{clouds}. However, our choices of
parameters listed in Tab.~\ref{tab:param} are such that they should
reproduce previous results from CCSN simulations. As these simulations
include not fully controlled ingredients, in particular initial data
for the progenitor stars, it is important to consider a general set of
profiles, unless future studies can bring techniques to determine
temperature and electron fraction profiles in CCSN from astrophysical
observable features. Last but not least, a limitation of our study is
the fact that we have considered the evolution of the PNS only during
the first second after its birth. In particular concerning the
problem~\ref{P1} about information of cold NS, the situation would of
course be quite different if longer evolution times were
considered. The particular value of 1~s post-bounce is justified by
CCSN simulations showing that during this period, a quasi-stationary
situation is achieved for the stalled accretion shock and
instability-induced perturbations of the PNS. It is not clear to what
extent strong-enough GW can be emitted later, and therefore whether it
is still possible to determine $M_{\rm PNS}$ and $R_{\rm PNS}$ at
later stages. This point would certainly be worth exploring more in
details, as later information would certainly be more decisive in
order to get insights on nuclear matter properties. A possibility in
this direction is evoked in~\cite{GalloRosso2018}, estimating the PNS
radius at about 6-10s after bounce from the neutrino signal. However,
for the moment the uncertainties entering the neutrino signal, in
particular those related to neutrino transport in the newborn hot PNS
and to neutrino-matter interactions, render a reliable extraction of the
PNS radius prohibitive.

\section*{Acknowledgements}

We warmly thank Pablo Cerd\'a-Dur\'an and Alejandro Torres-Forn\'e for
enlightening discussions, and Francesca Gulminelli for interesting
suggestions. The research leading to these results has received
funding from the PICS07889; it was also partially supported by the
Observatoire de Paris through the action f\'ed\'eratrice ``PhyFog''.



\bibliographystyle{mnras}
\bibliography{bib_pns} 





\bsp	
\label{lastpage}
\end{document}